# Gauge invariance and the Dirac free field energy

by

Dan Solomon

Rauland-Borg Corporation
3450 W. Oakton
Skokie, IL 60076

Phone: 847-324-8337
Email: dan.solomon@rauland.com

August 24, 2002

**Abstract**

Dirac field theory is assumed to be gauge invariant. However it is well known that a calculation of the polarization tensor yields a non-gauge invariant result. The reason for this has been shown to be due to the fact that for Dirac theory to be gauge invariant requires that there be no lower bound to the free field energy. Here an additional proof will be presented to show that this is indeed the case.

Dirac field theory is assumed to be gauge invariant [1]. However it is well known that when perturbation theory is used to calculate the polarization tensor the result is not gauge invariant (see Chapt. 14 of [2], Sect 22 of [3], Chapt. 5 of [4], and [5]). Non-gauge invariant terms must be isolated and removed from the calculation to obtain the correct gauge invariant result. This may involve some form of regularization, where other functions are introduced that have the correct behavior so that the non-gauge invariant terms are cancelled [4][5]. However, as has been pointed out by Pauli and Villars [5], there is no physical explanation for introducing these functions. They are merely mathematical devices used to force the desired (gauge invariant) result.

Now why does a theory that is presumably gauge invariant produce a non-gauge invariant result when calculations are performed? The reason that this problem occurs was recently investigated by this author[6]. It was shown that for Dirac theory to be gauge invariant requires that there be no lower bound to the free field energy. The free field energy is the energy of the quantum state in the absence of interactions. This result contradicts the generally held assumption that the vacuum state is the lower bound to the free field energy. In fact in standard quantum theory the vacuum state is defined in such a way that it is the state with the lowest free field energy. As is shown in [6] this will lead to a non-gauge invariant theory. To obtain a gauge invariant theory the definition of the vacuum state must be modified from its traditional definition to allow for the existence of quantum states with less free field energy than the vacuum. In [6] it was shown how this can be done and that when the "modified" vacuum state is used perturbation theory will be gauge invariant.

In this article, as in the previous work, it will be shown that the requirement of gauge invariance means that there can be no lower bound to the free field energy in Dirac theory. However, the proof presented here is significantly different from that given before. In [6] the proof relied on the dynamical equation governing the evolution of the state vector. Here we use the fact that a gauge transformation is a type of symmetry transformation. In the following we will work in the Schrodinger representation so that the time dependence of the system is assigned to the state vector and the operators are time independent. Also natural units are used so that $\hbar = c = 1$.

Consider the Dirac Hamiltonian in the presence of a static (time-independent) electric potential [6],

$$\hat{H}_D = \hat{H}_0 - \int \hat{\vec{J}}(\vec{x}) \cdot \vec{A}(\vec{x}) d\vec{x} + \int \hat{\rho}(\vec{x}) A_0(\vec{x}) d\vec{x} \tag{1}$$

In the above expression $(A_0, \vec{A})$ is the electric potential and is taken to be an unquantized, real valued quantity. The quantities $\hat{\vec{J}}(\vec{x})$ and $\hat{\rho}(\vec{x})$ are the current and charge operators, respectively, and $\hat{H}_0$ is the free field Hamiltonian, which is the Hamiltonian when the electric potential is zero. The time evolution of the state vector $|\Omega\rangle$ is given by,

$$i\frac{\partial |\Omega\rangle}{\partial t} = \hat{H}_D |\Omega\rangle \tag{2}$$

Assume that $|\Omega\rangle$ is normalized, i.e.,

$$\langle \Omega | \Omega \rangle = 1 \tag{3}$$

The energy in the presence of a static electric potential is,

$$E(|\Omega\rangle) = \langle\Omega|\hat{H}_D|\Omega\rangle \tag{4}$$

From (2) it can be shown that $E(|\Omega\rangle)$ is constant in time. The free field energy is the energy when the electric potential is zero.

It is generally assumed that there is a state vector $|0\rangle$, called the vacuum state, which has the following properties. First the current expectation value is zero, i.e.,

$$\langle 0|\hat{\vec{J}}|0\rangle = 0 \tag{5}$$

The second property is that the free field energy of $|0\rangle$ is zero and this represents a lower bound to the free field energy, i.e.,

$$\langle\Omega|\hat{H}_0|\Omega\rangle \geq \langle 0|\hat{H}_0|0\rangle = 0 \text{ for all } |\Omega\rangle \tag{6}$$

We will show that the above property is not consistent with the requirement that the theory be gauge invariant.

For a physical theory to be gauge invariant means that a change in the gauge cannot produce any change in any measurable results or observables. These include the current and charge expectation values and the energy difference between two different physical states. A change in the gauge is a change in the electric potential that does not produce a change in the electromagnetic field. For a static electromagnetic field the electric field $\vec{E}(\vec{x})$ and the magnetic field $\vec{B}(\vec{x})$ are given by,

$$\vec{E}(\vec{x}) = -\vec{\nabla}A_0(\vec{x}) \text{ and } \vec{B}(\vec{x}) = \vec{\nabla}\times\vec{A}(\vec{x}) \tag{7}$$

For such a static field a change in the gauge is given by,

$$\vec{A}(\vec{x}) \to \vec{A}'(\vec{x}) = \vec{A}(\vec{x}) + \vec{\nabla}\chi(\vec{x}) \tag{8}$$

where $\chi(\vec{x})$ is an arbitrary time-independent real valued function.

A gauge transformation is a type of symmetry transformation [2]. According to [7] a symmetry transformation is a change in our point of view that does not change the results of possible experiments. These include gauge transformations, Lorentz boosts, time or space translations, and rotations.

Let $\{|\Omega\rangle\}$ be a set of all possible state vectors. Consider two observers S and $S_g$ that are related by a symmetry transformation. For the observer S let $|\Omega_n\rangle \in \{|\Omega\rangle\}$ correspond to a given physical system. For the observer $S_g$ let the state vector that corresponds to the same physical system be given by $|\Omega_{n_g}\rangle \in \{|\Omega\rangle\}$. It is shown in [7] that the state vectors $|\Omega_n\rangle$ and $|\Omega_{n_g}\rangle$ are related by $|\Omega_{n_g}\rangle = \hat{U}|\Omega_n\rangle$ where $\hat{U}$ is a unitary operator that maps states in S into the corresponding states in $S_g$. Note that $\hat{U}|\Omega_n\rangle \in \{|\Omega\rangle\}$. Therefore $\hat{U}$ is a mapping of state vectors within the set $\{|\Omega\rangle\}$ and is a function of the parameters of the symmetry transformation.

Now let the symmetry transformation that relates S to $S_g$ be a gauge transformation. For S let the electric potential be $(A_0, \vec{A}) = 0$. Obviously the electromagnetic field is zero. For the observer $S_g$ let the electric potential be given by,

$$\vec{A}(\vec{x}) = \vec{\nabla}\chi(\vec{x}),\ A_0 = 0 \tag{9}$$

From (7) and (8) it is evident that the electromagnetic field for $S_g$ is also zero and that the electric potentials for S and $S_g$ are related by a gauge transformation. Therefore there is a unitary operator $\hat{U}_\chi$ that transforms a state vector $|\Omega_n\rangle$ in S into the state vector $|\Omega_{n_g}\rangle$ in $S_g$ according to,

$$\left|\Omega_{n_g}\right\rangle = \hat{U}_\chi \left|\Omega_n\right\rangle \tag{10}$$

The subscript $\chi$, in the above, denotes the fact that the unitary operator $\hat{U}_\chi$ is dependent on the parameters of the symmetry transformation which, in this case, is the function $\chi$.

From the requirements of gauge invariance we know that the state vectors $\left|\Omega_n\right\rangle$ and $\hat{U}_\chi \left|\Omega_n\right\rangle$ must satisfy the following relationship for the current expectation values,

$$\left\langle \Omega_n \left| \hat{\vec{J}} \right| \Omega_n \right\rangle = \left\langle \Omega_n \left| \hat{U}_\chi^\dagger \hat{\vec{J}} \hat{U}_\chi \right| \Omega_n \right\rangle \tag{11}$$

where by the usual rules we have used the fact that the dual of $\hat{U}_\chi \left|\Omega_n\right\rangle$ is $\left\langle \Omega_n \right| \hat{U}_\chi^\dagger$.

From (4) and (1) it is evident that the energy of a state vector $\left|\Omega\right\rangle$ in S is given by,

$$E\left(\left|\Omega\right\rangle, A^\mu = 0\right) = \left\langle \Omega \left| \hat{H}_0 \right| \Omega \right\rangle \tag{12}$$

Note that this is the same as the free field energy because the electric potential in S is zero. Use (9) in (1) along with (4) to show that the energy a state vector $\left|\Omega\right\rangle$ in $S_g$ is,

$$E\left(\left|\Omega\right\rangle, \vec{A} = \vec{\nabla}\chi\right) = \left\langle \Omega \left| \left( \hat{H}_0 - \int \hat{\vec{J}} \cdot \vec{\nabla}\chi \, d\vec{x} \right) \right| \Omega \right\rangle \tag{13}$$

From the principle of gauge invariance we have that the energy difference between the states $\left|\Omega_n\right\rangle$ and $\left|\Omega_m\right\rangle$, as viewed by observer S, must be the same as energy difference of the corresponding states, $\hat{U}_\chi \left|\Omega_n\right\rangle$ and $\hat{U}_\chi \left|\Omega_m\right\rangle$ as viewed by $S_g$, therefore using (12) and (13) we obtain,

$$\left\langle \Omega_n \left| \hat{H}_0 \right| \Omega_n \right\rangle - \left\langle \Omega_m \left| \hat{H}_0 \right| \Omega_m \right\rangle = \begin{pmatrix} \left\langle \Omega_n \left| \hat{U}_\chi^\dagger \left( \hat{H}_0 - \int \hat{\vec{J}} \cdot \vec{\nabla}\chi \, d\vec{x} \right) \hat{U}_\chi \right| \Omega_n \right\rangle \\ -\left\langle \Omega_m \left| \hat{U}_\chi^\dagger \left( \hat{H}_0 - \int \hat{\vec{J}} \cdot \vec{\nabla}\chi \, d\vec{x} \right) \hat{U}_\chi \right| \Omega_m \right\rangle \end{pmatrix} \tag{14}$$

Next consider the vacuum state $|0\rangle$ as defined by S. Assume that $|0\rangle$ satisfies the properties given in (5) and (6). From the previous discussion $\hat{U}_\chi|0\rangle$ is the state vector in $S_g$ that corresponds to $|0\rangle$ in S. According to the principle of gauge invariance if $|0\rangle$ is the lower bound to the energy as viewed from S then $\hat{U}_\chi|0\rangle$ is the lower bound to the energy as viewed by $S_g$. Therefore the relationship that must hold for the observer $S_g$ that corresponds to (6) is,

$$E\left(|\Omega\rangle, \vec{A}=\vec{\nabla}\chi\right) \geq E\left(\hat{U}_\chi|0\rangle, \vec{A}=\vec{\nabla}\chi\right) \text{ for all } |\Omega\rangle \tag{15}$$

From (13) the energy of the state $\hat{U}_\chi|0\rangle$ as seen by $S_g$ is,

$$E\left(\hat{U}_\chi|0\rangle, \vec{A}=\vec{\nabla}\chi\right) = \langle 0|\hat{U}_\chi^\dagger\left(\hat{H}_0 - \int\hat{\vec{J}}\cdot\vec{\nabla}\chi d\vec{x}\right)\hat{U}_\chi|0\rangle \tag{16}$$

Use (11) and (5) to yield,

$$\langle 0|\hat{U}_\chi^\dagger\hat{\vec{J}}\hat{U}_\chi|0\rangle = 0 \tag{17}$$

Use this in (16) to obtain,

$$E\left(\hat{U}_\chi|0\rangle, \vec{A}=\vec{\nabla}\chi\right) = \langle 0|\hat{U}_\chi^\dagger\hat{H}_0\hat{U}_\chi|0\rangle \tag{18}$$

Substitute this into (15) to obtain,

$$E\left(|\Omega\rangle, \vec{A}=\vec{\nabla}\chi\right) \geq \langle 0|\hat{U}_\chi^\dagger\hat{H}_0\hat{U}_\chi|0\rangle \text{ for all } |\Omega\rangle \tag{19}$$

Now what is $\langle 0|\hat{U}_\chi^\dagger\hat{H}_0\hat{U}_\chi|0\rangle$? To determine this we will substitute $|0\rangle$ for $|\Omega\rangle$ in (19). First refer to (13) and use (5) and (6) to obtain,

$$E\left(|0\rangle, \vec{A}=\vec{\nabla}\chi\right) = \langle 0|\left(\hat{H}_0 - \int\hat{\vec{J}}\cdot\vec{\nabla}\chi d\vec{x}\right)|0\rangle = 0 \tag{20}$$

Use this result in (15) to obtain,

$$E\left(|0\rangle, \vec{A} = \vec{\nabla}\chi\right) = 0 \geq \langle 0|\hat{U}_\chi^\dagger \hat{H}_0 \hat{U}_\chi|0\rangle \tag{21}$$

Next in (6) substitute $\hat{U}_\chi|0\rangle$ for $|\Omega\rangle$ to obtain,

$$\langle 0|\hat{U}_\chi^\dagger \hat{H}_0 \hat{U}_\chi|0\rangle \geq 0 \tag{22}$$

The only way that both (21) and (22) can be true is if,

$$\langle 0|\hat{U}_\chi^\dagger \hat{H}_0 \hat{U}_\chi|0\rangle = 0 \tag{23}$$

Next we will use the above results to show that there exist quantum states with a negative free field energy which contradicts (6). Refer to (14) and let $|\Omega_m\rangle = |0\rangle$ to obtain,

$$\langle \Omega_n|\hat{H}_0|\Omega_n\rangle - \langle 0|\hat{H}_0|0\rangle = \begin{pmatrix} \langle \Omega_n|\hat{U}_\chi^\dagger \left(\hat{H}_0 - \int \hat{\vec{J}}\cdot\vec{\nabla}\chi d\vec{x}\right)\hat{U}_\chi|\Omega_n\rangle \\ -\langle 0|\hat{U}_\chi^\dagger \left(\hat{H}_0 - \int \hat{\vec{J}}\cdot\vec{\nabla}\chi d\vec{x}\right)\hat{U}_\chi|0\rangle \end{pmatrix} \tag{24}$$

Next use (23) and (6) in the above to obtain,

$$\langle \Omega_n|\hat{H}_0|\Omega_n\rangle = \begin{pmatrix} \langle \Omega_n|\hat{U}_\chi^\dagger \left(\hat{H}_0 - \int \hat{\vec{J}}\cdot\vec{\nabla}\chi d\vec{x}\right)\hat{U}_\chi|\Omega_n\rangle \\ +\langle 0|\hat{U}_\chi^\dagger \left(\int \hat{\vec{J}}\cdot\vec{\nabla}\chi d\vec{x}\right)\hat{U}_\chi|0\rangle \end{pmatrix} \tag{25}$$

Refer to (11) and (5) to write,

$$\langle \Omega_n|\hat{U}_\chi^\dagger \hat{\vec{J}} \hat{U}_\chi|\Omega_n\rangle = \langle \Omega_n|\hat{\vec{J}}|\Omega_n\rangle \text{ and } \langle 0|\hat{U}_\chi^\dagger \hat{\vec{J}} \hat{U}_\chi|0\rangle = \langle 0|\hat{\vec{J}}|0\rangle = 0 \tag{26}$$

Use this in (25) to obtain,

$$\langle \Omega_n|\hat{H}_0|\Omega_n\rangle = \langle \Omega_n|\hat{U}_\chi^\dagger \hat{H}_0 \hat{U}_\chi|\Omega_n\rangle - \int \langle \Omega_n|\hat{\vec{J}}|\Omega_n\rangle \cdot \vec{\nabla}\chi d\vec{x} \tag{27}$$

Rearrange terms to obtain,

$$\langle \Omega_n|\hat{U}_\chi^\dagger \hat{H}_0 \hat{U}_\chi|\Omega_n\rangle = \langle \Omega_n|\hat{H}_0|\Omega_n\rangle + \int \langle \Omega_n|\hat{\vec{J}}|\Omega_n\rangle \cdot \vec{\nabla}\chi d\vec{x} \tag{28}$$

Integrate by parts and assume reasonable boundary conditions to obtain,

$$\langle\Omega_n|\hat{U}_\chi^\dagger\hat{H}_0\hat{U}_\chi|\Omega_n\rangle = \langle\Omega_n|\hat{H}_0|\Omega_n\rangle - \int \chi\vec{\nabla}\cdot\langle\Omega_n|\hat{\vec{J}}|\Omega_n\rangle d\vec{x} \qquad (29)$$

The quantities $\langle\Omega_n|\hat{H}_0|\Omega_n\rangle$ and $\langle\Omega_n|\hat{\vec{J}}|\Omega_n\rangle$, in the above equation, are independent of the function $\chi(\vec{x})$. Assume $|\Omega_n\rangle$ is chosen so that $\langle\Omega_n|\hat{\vec{J}}|\Omega_n\rangle \neq 0$. Then it is always possible to find a $\chi(\vec{x})$ which makes $\langle\Omega_n|\hat{U}_\chi^\dagger\hat{H}_0\hat{U}_\chi|\Omega_n\rangle$ a negative number with an arbitrarily large magnitude. For example let,

$$\chi(\vec{x}) = f\vec{\nabla}\cdot\langle\Omega_n|\hat{\vec{J}}|\Omega_n\rangle \qquad (30)$$

Use this in (29) to yield,

$$\langle\Omega_n|\hat{U}_\chi^\dagger\hat{H}_0\hat{U}_\chi|\Omega_n\rangle = \langle\Omega_n|\hat{H}_0|\Omega_n\rangle - f\int\left(\vec{\nabla}\cdot\langle\Omega_n|\hat{\vec{J}}|\Omega_n\rangle\right)^2 d\vec{x} \qquad (31)$$

The integral in the above equation is always positive. Therefore as $f \to \infty$ the quantity $\langle\Omega_n|\hat{U}_\chi^\dagger\hat{H}_0\hat{U}_\chi|\Omega_n\rangle$ becomes a negative number with an arbitrarily large magnitude.

In conclusion, it has been shown that if Dirac theory is gauge invariant, equation (6) cannot be true for all possible state vectors $|\Omega\rangle$. This is because, as shown above, there exists a state vector $|\Omega_{n_g}\rangle = \hat{U}_\chi|\Omega_n\rangle$ whose free field energy $\langle\Omega_{n_g}|\hat{H}_0|\Omega_{n_g}\rangle = \langle\Omega_n|\hat{U}_\chi^\dagger\hat{H}_0\hat{U}_\chi|\Omega_n\rangle$ is a negative number with an arbitrarily large magnitude. It is shown in [6] that this is the reason why perturbation calculations do not yield gauge invariant results.